\documentclass[iop,superscriptaddress]{emulateapj}

\usepackage{graphicx}
\usepackage{dcolumn}
\usepackage{amssymb} 
\usepackage{ctable}
\usepackage{bm}
\usepackage{epsfig}
\usepackage{epstopdf}
\usepackage{epsf,color}
\usepackage{natbib}

\newcommand{\cmnt}[1]{}

\begin{document}

\title{Probing Gravity at Large Scales through CMB Lensing}

\author{Anthony R. Pullen, Shadab Alam, and Shirley Ho}
\affiliation{Department of Physics, Carnegie Mellon University, 5000 Forbes Ave, Pittsburgh, PA, 15213, U.S.A.}

\email{apullen@andrew.cmu.edu}

\begin{abstract}
We describe a methodology to probe gravity with the cosmic microwave background (CMB) lensing convergence $\kappa$, specifically by measuring $E_G$, the ratio of the Laplacian of the gravitational scalar potential difference with the velocity divergence.  Using CMB lensing instead of galaxy-galaxy lensing avoids intrinsic alignments while also lacking a hard limit on the lens redshift and significant uncertainties in the source plane.  We model $E_G$ for general relativity and modified gravity, finding that $E_G$ for $f(R)$ gravity should be scale-dependent due to the scale-dependence of the growth rate $f$.  Next, we construct an estimator for $E_G$ in terms of the galaxy-CMB lensing and galaxy clustering angular power spectra, along with the RSD parameter $\beta$.  We also forecast statistical errors for $E_G$ from the current \emph{Planck} CMB lensing map and the spectroscopic galaxy and quasar samples from the Sloan Digital Sky Survey Data Release 11, being 9\% with galaxies and 8\% when quasars are included.  We also find that upcoming spectroscopic and photometric surveys, combined with the final \emph{Planck} lensing map, can measure precisely the redshift- and scale-dependence of $E_G$ out to redshifts $z=2$ and higher, with photometric surveys having an advantage due to their high number densities.   Advanced ACTPol's lensing map will increase the $E_G$ sensitivity even further.  Finally, we find that Advanced ACTPol cross-correlated with spectroscopic (photometric) surveys can differentiate between general relativity and $f(R)$ gravity at the level of $3\sigma$ ($13\sigma$).  Performing a $<1$\% measurement of $E_G$ requires a 10\% precision in $\beta$ from \emph{Euclid} or LSST, currently achievable with a spectroscopic survey but difficult with only a photometric survey.
\end{abstract}
\maketitle

\section{Introduction}
The discovery of cosmic acceleration \citep{1998AJ....116.1009R,1999ApJ...517..565P} has inspired numerous theoretical explanations for its existence. On one hand, the acceleration can be caused by a new, unknown force that exhibits negative pressure called \emph{dark energy} \citep{2003RvMP...75..559P}.  The cosmological constant, a special case of dark energy, is a major component of the $\Lambda$CDM framework that explains the expansion history and the growth history of the Universe \citep{2014MNRAS.441...24A} and the cosmic microwave background (CMB) \citep{2013ApJS..208...20B,2013arXiv1303.5076P}.  On the other hand, it is possible that the dynamics of gravity deviate from general relativity (GR) on cosmological scales \citep{2000PhLB..485..208D,2004PhRvD..70d3528C}, a concept called \emph{modified gravity}.  In observations of the universe's expansion history, these two effects are degenerate with one another.  However, we expect the growth of structure to differ between dark energy and modified gravity models such that the degeneracy is broken.

An observable that probes the expansion history and growth of structure simultaneously is $E_G$ \citep{2007PhRvL..99n1302Z}, which is related to the ratio of the Laplacian of the difference between the two scalar potentials $\nabla^2(\psi-\phi)$ to the peculiar velocity perturbation field $\theta$.  The value of $E_G$ depends on how gravity behaves on large scales.  Traditionally, when measuring $E_G$, $\nabla^2(\psi-\phi)$ is probed by a galaxy lensing correlation with a tracer of large-scale structure (LSS), while the peculiar velocity field is probed by either a galaxy-velocity cross-correlation or, equivalently, a galaxy autocorrelation times the redshift-space distortion (RSD) parameter $\beta=f/b$, where $f$ is the growth rate and $b$ is the clustering bias of the galaxies.  A major advantage of $E_G$ over other observables is that it is independent of clustering bias on linear scales, reducing the model uncertainty.  $E_G$ was first measured in \citet{2010Natur.464..256R}, using galaxy-galaxy lensing exhibited by Sloan Digital Sky Survey (SDSS) \citep{2000AJ....120.1579Y} Luminous Red Galaxies (LRGs) \citep{2001AJ....122.2267E} to find $E_G(z=0.32)=0.392\pm0.065$, consistent with $\Lambda$CDM.

In this analysis we assess the possibility of using CMB lensing \citep{1987A&A...184....1B,1988A&A...206..199L,1989MNRAS.239..195C} cross correlated with galaxies \citep{2004PhRvD..70j3501H,2007PhRvD..76d3510S,2008PhRvD..78d3520H,2014A&A...571A..17P} as a probe of $\nabla^2(\psi-\phi)$ instead of the traditional method of using galaxy lensing.  One advantage of using CMB lensing over galaxy lensing is that the CMB lensing kernel is very broad over redshift, allowing probes of $E_G$ at much higher redshifts than with galaxy lensing. At these higher redshifts, CMB lensing also has the added bonus of probing more linear scales, reducing systematic effects due to nonlinear clustering.  Also, since the CMB propagates throughout all of space, all of LSS sampled by the survey lenses the CMB, allowing us to measure the lensing part of $E_G$ at much higher redshifts.  We also do not have to worry about complex astrophysical uncertainties of the source galaxies, \emph{i.e.} intrinsic alignments, since the CMB is simple.  Finally, we know the CMB redshift, so we can avoid determining the photometric redshift distribution of the sources.  We construct an estimator for $E_G$ in terms of the angular cross-power spectrum between the CMB lensing convergence $\kappa$ and galaxies, the angular auto-power spectrum of the same galaxies, and the RSD parameter $\beta$.

Next, we derive $E_G$ for general relativity (GR), as well as modified gravity using the $\mu\gamma$ formalism from \citet{2011JCAP...08..005H}.  While $E_G(z)$ can be written as $\Omega_{m,0}/f(z)$ for $\Lambda$CDM, where $\Omega_{m,0}$ is the matter density today relative to the critical density and $f(z)$ is the growth rate at redshift $z$, $E_G$ for modified gravity models is expected to differ from this value.  We also found that $E_G$ for $f(R)$ gravity \citep{2004PhRvD..70d3528C} and chameleon gravity \citep{2004PhRvL..93q1104K} can exhibit scale-dependence through $f$, which could potentially help differentiate between these gravity models and GR.

Next, we consider the prospect of measuring $E_G$ with CMB lensing by forecasting errors for an $E_G$ measurement using the current \emph{Planck} CMB lensing map \citep{2014A&A...571A..17P} along with the CMASS galaxy ($z=0.57$), LOWZ galaxy ($z=0.32$), and BOSS quasar \citep{2014A&A...563A..54P} spectroscopic samples from Data Release 11 (DR11) \citep{2014MNRAS.441...24A} of the SDSS-III \citep{2011AJ....142...72E} Baryon Oscillations Spectroscopic Survey (BOSS) \citep{2013AJ....145...10D}.  We also consider possibilities with upcoming surveys.  First, we consider the final \emph{Planck} lensing map cross-correlated with spectroscopic surveys, specifically the Dark Energy Spectroscopic Instrument (DESI) \citep{2013arXiv1308.0847L}, \emph{Euclid} \citep{2011arXiv1110.3193L}, and the Wide Field InfraRed Survey Telescope (WFIRST) \citep{2013arXiv1305.5422S}. We find, however, that an $E_G$ measurement using the \emph{Planck} CMB lensing map and upcoming spectroscopic surveys is not sensitive enough to differentiate between GR and $f(R)$ gravity at current limits, and that the new limits on chameleon gravity would be modest. We find that spectroscopic surveys with Advanced ACTPol can differentiate between GR and $f(R)$ gravity at the level of $3\sigma$, with higher significances for the chameleon gravity model.

We also consider the CMB lensing maps cross-correlated with photometric surveys, specifically the Dark Energy Survey (DES) \citep{2005astro.ph.10346T}, the Large Synoptic Survey Telescope (LSST) \citep{2009arXiv0912.0201L}, and \emph{Euclid}.  We find that for the scales which $E_G$ dominates, being small but still only quasi-linear, the lensing measurement dominates the error in $E_G$ as opposed to the RSD.  Thus, reducing the shot noise by increasing the survey number density is more important than having more precise redshifts for RSD.  We find DES is comparable in power to DESI, and LSST and photometric \emph{Euclid} can discriminate between GR and $f(R)$ gravity at very high significance using lensing from Adv.~ACTPol.  However, this will require an RSD precision on the order of 10\%, which is difficult but possible for \emph{Euclid} and LSST. Photometric surveys combined with CMB lensing experiments can produce significant constraints to $E_G$ that could help uncover the true nature of gravity.

The plan of the paper is as follows: in Section \ref{S:theory}, we write the theoretical $E_G$ for modified gravity, while Section \ref{S:estim} gives the estimator for $E_G$ in terms of CMB lensing.  In Section \ref{S:forecasts} we construct forecasts for various experiment configurations, and in Section \ref{S:conclude} we present conclusions.  We assume the combined CMASS/\emph{Planck} cosmology \citep{2013arXiv1303.5076P,2014MNRAS.441...24A} with $\Omega_mh^2=0.1418$, $h=0.676$, $\Omega_bh^2=0.0224$, $n_s=0.96$, and $\sigma_8=0.8$.

\section{Theory}\label{S:theory}

We begin with the definition of $E_G$ in Fourier space from \citet{2007PhRvL..99n1302Z}, given by
\begin{eqnarray}\label{E:eg}
E_G(k,z)&=&\frac{c^2[\nabla^2(\psi-\phi)]_k}{3H_0^2(1+z)\theta(k)}\nonumber\\
&=&\frac{c^2k^2(\phi-\psi)}{3H_0^2(1+z)\theta(k)}\, ,
\end{eqnarray}
where $H_0$ is the Hubble parameter today and $\theta(k)$ is the perturbation in the matter velocity field.  Assuming a flat universe described by a Friedmann-Robertson-Walker (FRW) metric with perturbation fields $\psi$ in the time component and $\phi$ in the spatial component, as well as negligible anisotropic stress and non-relativistic matter species, the time-time and momentum Einstein field equations in general relativity (GR) can be written in Fourier space as \citep{2011JCAP...08..005H}
\begin{eqnarray}
k^2\psi&=&-4\pi G a^2\rho(a)\delta\nonumber\\
\phi&=&-\psi\, ,
\end{eqnarray}
where $a$ is the scale factor, $\rho$ is the background matter density, and $\delta$ is the matter density perturbation.  These equations are generalized to a modified gravity (MG) model such that
\begin{eqnarray}\label{E:mg}
k^2\psi&=&-4\pi G a^2\mu(k,a)\rho(a)\delta\nonumber\\
\phi&=&-\gamma(k,a)\psi\, ,
\end{eqnarray}
where $\mu(k,a)$ and $\gamma(k,a)$ are arbitrary functions of $k$ and $a$, and $\mu=\gamma=1$ for GR. 

Using this formalism, we can write the numerator of $E_G$ in Eq.~\ref{E:eg} as
\begin{eqnarray}
k^2(\phi-\psi)&=&-k^2[1+\gamma(k,a)]\psi \nonumber\\
&=&4\pi G a^2\rho(a)\mu(k,a)[1+\gamma(k,a)]\delta\, .
\end{eqnarray}
Substituting $\rho(a)=\rho_0a^{-3}$ and $\Omega_{m,0}=8\pi G\rho_0/3H_0^2$, we find
\begin{eqnarray} \label{E:eg1}
k^2(\phi-\psi)=\frac{3}{2}H_0^2\Omega_{m,0}(1+z)\mu(k,a)[\gamma(k,a)+1]\delta
\end{eqnarray}
The velocity perturbation $\theta$ can be written as $\theta=f\delta$ on linear scales.  Combining this expression and Eq.~\ref{E:eg1} gives $E_G$ from Eq.~\ref{E:eg} as
\begin{eqnarray}
E_G(k,z)=\frac{\Omega_{m,0}\mu(k,a)[\gamma(k,a)+1]}{2f}\, .
\end{eqnarray}
This expression gives us the correct value in the GR limit, namely $E_G=\Omega_{m,0}/f(z)$.

For $f(R)$ gravity \citep{2004PhRvD..70d3528C} using the parametrization in \citet{2007PhRvD..75d4004S}, $\mu$ and $\gamma$ are given by \citep{2007PhRvD..76b3514T,2011JCAP...08..005H}
\begin{eqnarray}
\mu^{\rm fR}(k,a,)&=&\frac{1}{1-B_0a^{s-1}/6}\left[\frac{1+(2/3)B_0\bar{k}^2a^s}{1+(1/2)B_0\bar{k}^2a^s}\right]\nonumber\\
\gamma^{\rm fR}(k,a)&=&\frac{1+(1/3)B_0\bar{k}^2a^s}{1+(2/3)B_0\bar{k}^2a^s}\, ,
\end{eqnarray}
where $\bar{k}=k[2997.9\,{\rm Mpc}/h]$, $h=H_0/[100\,{\rm km/s/Mpc}]$, $s$=4 for models that follow the $\Lambda CDM$ expansion history, and $B_0$ is a free parameter which is related to the Compton wavelength of an extra scalar degree of freedom and is proportional to the curvature of $f(R)$ today.  Current measurements limit $B_0<5.6\times10^{-5}$ (1$\sigma$) \citep{2014arXiv1411.4353X,2014arXiv1406.3347B}.  For this gravity model, $E_G$ is given by
\begin{eqnarray}
E_G^{\rm BZ}(k,z)=\frac{1}{1-B_0a^{s-1}/6}\frac{\Omega_{m,0}}{f^{\rm BZ}(k,z)}\, ,
\end{eqnarray}
where $f^{\rm BZ}(k,z)$ is the BZ growth rate, which is scale-dependent since $\mu$ is scale-dependent \citep{2011JCAP...08..005H}.

Chameleon gravity \citep{2004PhRvL..93q1104K} is a Yukawa-type dark matter interaction equivalent to a class of scalar-tensor theories with a scalar field non-minimally coupled to the metric.  For chameleon gravity, using the Bertschinger and Zukin (BZ) parametrization \citep{2008PhRvD..78b4015B}, $\mu$ and $\gamma$ are given by \citep{2011JCAP...08..005H}
\begin{eqnarray}
\mu^{\rm Ch}(k,a)&=&\frac{1+\beta_1\lambda_1^2k^2a^s}{1+\lambda_1^2k^2a^s}\nonumber\\
\gamma^{\rm Ch}(k,a)&=&\frac{1+\beta_2\lambda_2^2k^2a^s}{1+\lambda_2^2k^2a^s}\nonumber\\
\lambda_2^2&=&\beta_1\lambda_1^2\nonumber\\
\beta_2&=&\frac{2}{\beta_1}-1\, ,
\end{eqnarray}
where the typical ranges for $\beta_1$ and $s$ are $0<\beta_1<2$ and $1\leq s\leq 4$.  We will relate $\lambda_1$ to $B_0$, a parameter similar to that for $f(R)$ gravity but much less constrained due to the extra degree of freedom in the model.  The typical range for $B_0$ in this model is [0,1] and is related to $\lambda_1$ by
\begin{eqnarray}
B_0=\frac{2\lambda_1^2H_0^2}{c^2}\, .
\end{eqnarray}

\begin{figure}
\begin{center}
\includegraphics[width=0.5\textwidth]{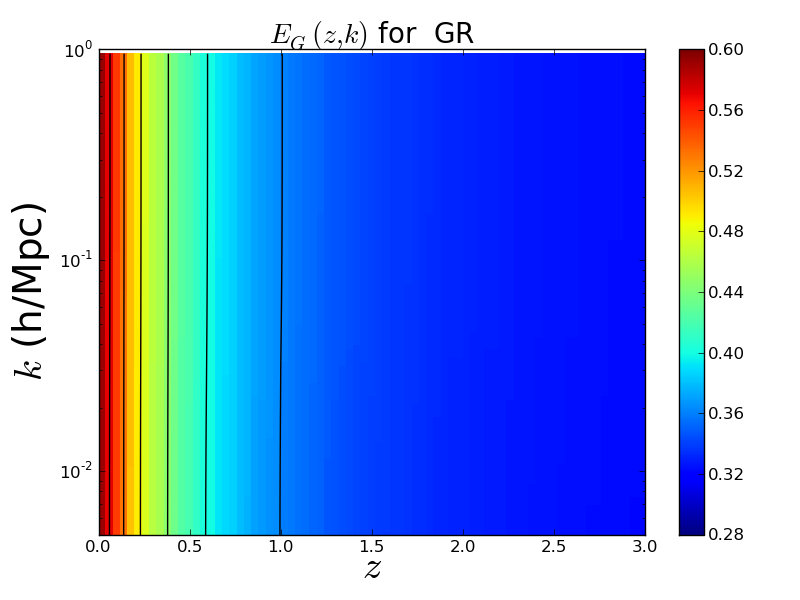}
\includegraphics[width=0.5\textwidth]{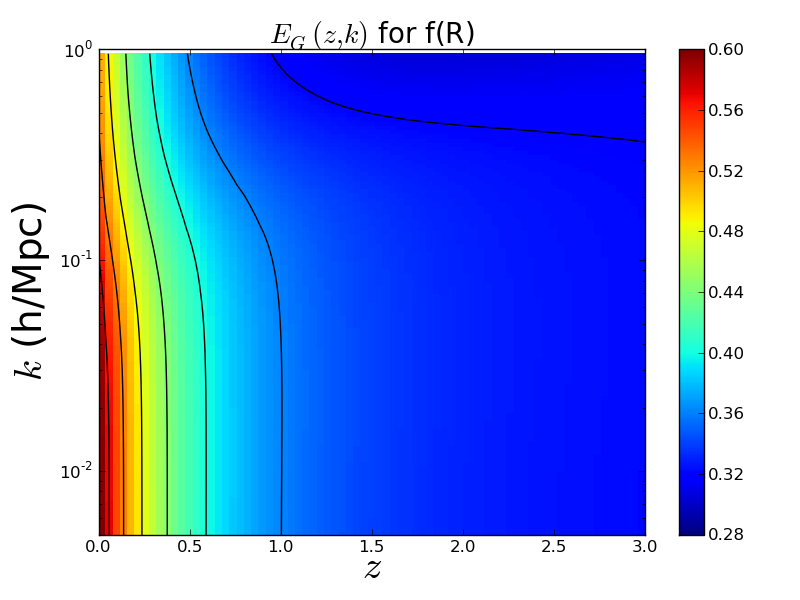}
\includegraphics[width=0.5\textwidth]{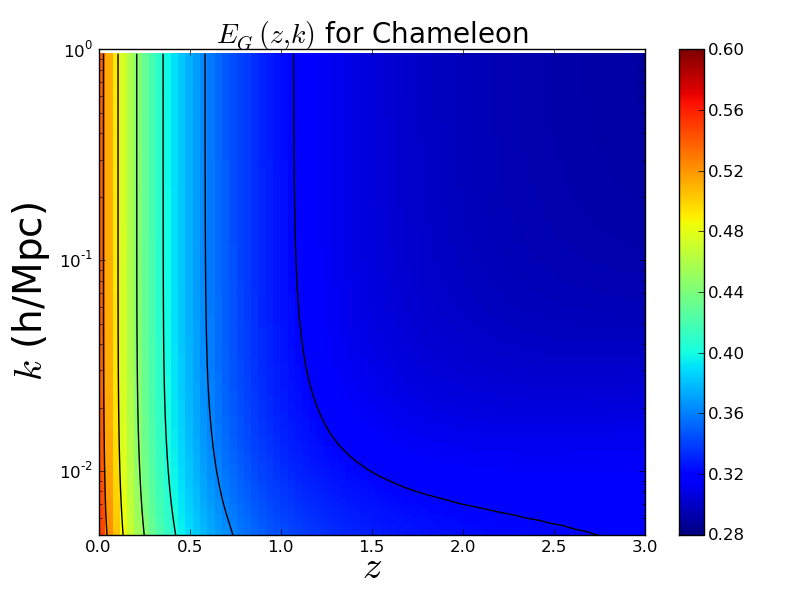}
\caption{\label{F:e.g.} $E_G$ as a function of redshift $z$ and wavenumber $k$ for GR (top), $f(R)$ gravity ($B_0=5.6\times10^{-5}$) (middle), and chameleon gravity ($\beta_1=1.2$, $B_0=0.4$, $s=4$) (bottom).  $E_G$ for $f(R)$ gravity exhibits strong scale dependence, decreasing by $\sim10\%$ at small scales.}
\end{center}
\end{figure}

We calculate $E_G(k,z)$ for GR, $f(R)$ gravity, and chameleon gravity using MGCAMB \citep{2000ApJ...538..473L,2011JCAP...08..005H}, which we plot in Fig.~\ref{F:e.g.}.  In GR, $E_G$ is scale-independent.  However, in $f(R)$ gravity $E_G$ decreases at small scales by $\sim10\%$, with the decrease being more pronounced at higher redshifts and smaller scales.  On the other hand, chameleon gravity shifts $E_G$ to lower values with respect to GR.   These results show that the ability to measure the scale-dependence of $E_G$ will be advantageous for constraining MG models, particularly for $f(R)$ gravity.

\section{Estimator} \label{S:estim}

Here we derive an estimator for $E_G$ in terms of the galaxy-convergence cross-power spectrum $C_\ell^{g\kappa}$, the galaxy auto-power spectrum $C_\ell^{gg}$, and the RSD parameter $\beta$.  The estimator is similar to the expression in \citet{2010Natur.464..256R} using the galaxy-galaxy lensing cross-power spectrum.

Starting with Eq.~\ref{E:eg}, $E_G$ can be estimated in terms of power spectra as
\begin{eqnarray}
\hat{E}_G(k,z)=\frac{c^2\hat{P}_{\nabla^2(\psi-\phi)g}(k)}{3H_0^2(1+z)\hat{P}_{\theta g}(k)}\, ,
\end{eqnarray}
where $P_{\nabla^2(\psi-\phi)g}$ is the galaxy-$\nabla^2(\psi-\phi)$ cross-power spectrum and $P_{\theta g}$ is the galaxy-peculiar velocity cross-power spectrum.  Projecting 3D power spectra into angular quantities, we can estimate $E_G$ as
\begin{eqnarray}
\hat{E}_G(\ell,\bar{z})=\frac{c^2\hat{C}_\ell^{\kappa g}}{3H_0^2\hat{C}_\ell^{
\theta g}}\, .
\end{eqnarray}
$C_\ell^{\kappa g}$ is the galaxy-convergence angular cross-power spectrum, given on small scales using the Limber approximation\footnote{Since we are mainly interested in small scales ($\ell\gtrsim100$), the Limber approximation is valid for most of the cases we consider (see Sec.~\ref{S:uss})} by
\begin{eqnarray}
C_\ell^{\kappa g}=\frac{1}{2}\int_{\chi_1}^{\chi_2}d\chi\,W(\chi)f_g(\chi)\chi^{-2}P_{\nabla^2(\psi-\phi)g}\left(\frac{\ell}{\chi},z\right)\, ,
\end{eqnarray}
where $f_g$ is the galaxy redshift distribution, $\chi_1$ and $\chi_2$ are the limits of the redshift distribution, and $W(\chi)=\chi(1-\chi/\chi_{\rm CMB})$ is the CMB lensing kernel.  In order to match the kernel in the galaxy-convergence power spectrum, we define $C_\ell^{\theta g}$, the velocity-galaxy angular cross-power spectrum, to be
\begin{eqnarray}\label{E:ctg}
C_\ell^{\theta g}&=&\frac{1}{2}\int_{\chi_1}^{\chi_2}d\chi\,W(\chi)f_g(\chi)\chi^{-2}(1+z)P_{\theta g}\left(\frac{\ell}{\chi},z\right)\, .
\end{eqnarray}
This cross-power spectrum is a construct used to measure $E_G$ without multiplicative bias, and is not equivalent to the RSD angular power spectrum derived in \citet{2007MNRAS.378..852P}.

In our analysis, as in \citet{2010Natur.464..256R}, we assume that the RSD parameter $\beta$ will be measured separately, and we approximate the lensing kernel and $\beta$ as constants over redshift within the integral, and we approximate $f_g(\chi)\simeq f_g^2(\chi)/f_g(\bar{\chi})$ within the redshift range.  This approximation works well when the galaxy redshift distribution is highly peaked, which is especially true when we use small redshift bins in forecasts for upcoming surveys.  Also, we assume from linear theory $\theta=f\delta$.  In that case, Eq.~\ref{E:ctg} can be written as
\begin{eqnarray} \label{E:clapprox}
C_\ell^{\theta g}&\simeq&\frac{W(\bar{\chi})(1+\bar{z})}{2f_g(\bar{\chi})}\int_{\chi_1}^{\chi_2}d\chi\,f_g^2(\chi)\chi^{-2}\beta (z)P_{gg}\left(\frac{\ell}{\chi},z\right)\nonumber\\
&\simeq&\frac{W(\bar{\chi})\beta(\bar{z})(1+\bar{z})}{2f_g(\bar{\chi})}C_\ell^{gg}\, ,
\end{eqnarray}
where $C_\ell^{gg}$ is the galaxy angular auto-power spectrum.  Note that $\beta$ could exhibit scale-dependence due to modified gravity through the growth rate.   Thus, $E_G$ in this case can be written as
\begin{eqnarray}
\hat{E}_G(\ell,\bar{z})=\frac{2c^2\hat{C}_\ell^{\kappa g}}{3H_0^2(1+\bar{z})W(\bar{\chi})\Delta\chi\beta(\bar{z})\hat{C}_\ell^{gg}}\, ,
\end{eqnarray}
and we can write the error of $E_G$ in terms of the errors of $\beta$ and $C_\ell^{gg}$ as
\begin{eqnarray}\label{E:segb}
\frac{\sigma^2[E_G(\ell,\bar{z})]}{E_G^2}= \left[\left(\frac{\sigma(C_\ell^{\kappa g})}{C_\ell^{\kappa g}}\right)^2+\left(\frac{\sigma(\beta)}{\beta}\right)^2+\left(\frac{\sigma(C_\ell^{gg})}{C_\ell^{gg}}\right)^2\right]\, ,
\end{eqnarray}
which is strictly true if $\beta$ and $C_\ell^{gg}$ are measured from separate surveys.  This approximation does underestimate the error if $\beta$ and $C_\ell^{gg}$ are measured from the same survey.  However, it should not be much of a concern for us.  For current surveys and upcoming spectroscopic surveys, the error is mainly dominated by the CMB lensing, which is a measure independent of $\beta$ and $C_\ell^{gg}$.  For example, switching between independent to dependent measures of $\beta$ and $C_\ell^{gg}$ for BOSS reduces the signal-to-noise ratio for $E_G$ by 1\%.  For upcoming photometric surveys when the lensing error is competitive with the error in $\beta$, the error in the galaxy clustering due to shot noise should be small enough such that it can be neglected, making the $\beta C_\ell^{gg}$ error independent of this concern.  For example, switching between independent to dependent measures of $\beta$ and $C_\ell^{gg}$ for \emph{Euclid} photometric survey with Adv.~ACTPol reduces the signal-to-noise ratio for $E_G$ by 5\%.  The uncertainty in $C_\ell^{\kappa g}$ can be written as
\begin{eqnarray}
\sigma^2(C_\ell^{\kappa g}) = \frac{(C_\ell^{\kappa g})^2+(C_\ell^{\kappa\kappa}+N_\ell^{\kappa\kappa})(C_\ell^{gg}+N^{gg})}{(2\ell+1)f_{\rm sky}}\, ,
\end{eqnarray}
where $C_\ell^{\kappa\kappa}$, the convergence auto-power spectrum, is computed from CAMB \citep{2000ApJ...538..473L}, $N_\ell^{\kappa\kappa}$ is the noise in the convergence power spectrum computed using the formalism in \citet{2002ApJ...574..566H}, and $N^{gg}$ is the shot noise.

A precise measurement of $E_G$ will be slightly biased from the true value due to several reasons similar to those outlined in the Appendix of \citet{2010Natur.464..256R}.  For one, galaxy redshift distributions can exhibit a spread and skewness, causing the approximation in Eq.~\ref{E:clapprox} to break down.  However, we can characterize this deviation analytically as a function of angular scale.  Also, in order to extend our measurement of $E_G$ to small scales, we must correct for the scale-dependence of the bias due to clustering at nonlinear scales.  We may also need to consider scale-dependent $\beta$ due to nonlinear density and velocity perturbations, although velocity perturbations tend to stay linear at smaller scales than for density perturbations.  In addition, constraints on modified gravity parameters will have to account for the scale dependence of $\beta$ due to the growth rate.  We expect these effects to be small and will neglect them in our forecasts.

\section{Forecasts} \label{S:forecasts}

In this section we will predict the ability of current and future surveys to measure $E_G$ and differentiate between GR and MG models.  In all our forecasts we will assume GR when calculating uncertainties.  We describe the sensitivity of the measured $E_G$ with the signal-to-noise ratio (SNR) of $E_G$ marginalized over angular scale and redshift, given by
\begin{eqnarray}
{\rm SNR}^2(E_G)=\sum_{\ell,z_i}\frac{[E_G^{\rm GR}(z_i)]^2}{\sigma^2[E_G(\ell,z_i)]}\, ,
\end{eqnarray}
where $z_i$ denotes redshift bins.  We also calculate the $\chi^2$ value between GR and MG models to determine if a particular $E_G$ measurement could distinguish between GR and a given MG model.  We write $\chi^2$ as
\begin{eqnarray}
\chi^2(E_G)=\sum_{\ell,z_i}\frac{[E_G^{\rm MG}(\ell,z_i)-E_G^{\rm GR}(z_i)]^2}{\sigma^2[E_G(\ell,z_i)]}\, ,
\end{eqnarray}
where $E_G^{\rm MG}(\ell,z_i)$ is the $E_G$ estimate for a given MG model, which is generally $\ell$-dependent.  This $\chi^2$ is related to the likelihood ratio between GR and modified gravity, assuming modified gravity is the true theory, and $\chi^2\gg1$ means the two theories can be differentiated at high significance. Throughout the section we quote $\chi_{\rm rms}=\sqrt{\chi^2}$.  Note that for the following limits, $f(R)$ gravity is set to its upper limit value $B_0=5.6\times10^{-5}$, and that chameleon gravity's base set of parameters is $B_0=0.4$, $\beta_1=1.2$, and $s=4$, unless otherwise stated.

\subsection{Current Surveys}

We begin with forecasts of $E_G$ measurements from the publicly available \emph{Planck} CMB lensing map \citep{2014A&A...571A..17P} and the CMASS and LOWZ spectroscopic galaxy samples from BOSS DR11 \citep{2014MNRAS.441...24A}, as well as the spectroscopic quasar (QSO) sample \citep{2014A&A...563A..54P} from DR11.  We use the noise estimate given in the public \emph{Planck} CMB lensing map.   The total number of galaxies (or quasars) within each sample along with the survey area are listed in Table \ref{T:galsurveys}.  For CMASS, we use the measurements of $b\sigma_8$ and $f\sigma_8$ from \citet{2014MNRAS.439.3504S} to set $b$[CMASS]=2.16 and $\sigma(\beta)/\beta\sim$ 10\% for the entire redshift range.  The corresponding values for the LOWZ sample have not been measured model-independently for DR11, so we assume a 10\% measurement of $\beta$ for the entire redshift range and, as in \citet{2014MNRAS.440.2222T}, we assume $b$[LOWZ]=1.85. We use Eq.~\ref{E:segb} to calculate the $E_G$ uncertainty for these samples.  For the BOSS quasar sample, we use the BOSS DR9 bias measurements from \citet{2012MNRAS.424..933W}, assuming the average value $b_{\rm QSO}=3.83$.  We also assume a 15\% $\beta$ measurement in each of two redshift bins for the BOSS quasars.  This is a bit optimistic, considering there are systematic issues on linear scales with measuring RSD from BOSS quasars.  However, we confirm that even a measurement error of 100\% [$\sigma(\beta)/\beta=1$] only increases the errors on $E_G$ by 5\%.

\begin{table}
\begin{center}
\caption{\label{T:galsurveys} Properties of the various spectroscopic surveys considered in our analysis.} 
\begin{tabular}{c|ccc}
\hline
Survey&$z$&Area (deg$^2$)&$N_{\rm gal}$\\
\hline
BOSS CMASS&0.43--0.7&7900&704,000\\
BOSS LOWZ&0.15--0.43&6900&306,000\\
BOSS QSOs&2.1--3.5&10,200&175,000\\
DESI ELGs&0.6--1.7&14,000&$1.8\times10^7$\\
DESI LRGs&0.6--1.2&14,000&$4.1\times10^6$\\
DESI QSOs&0.6--1.9&14,000&$1.9\times10^6$\\
\emph{Euclid}&0.5--2.0&20,000&$1.7\times10^7$\\
WFIRST&1.05-2.9&2000&$3.0\times10^7$\\
\hline
\end{tabular}\end{center}
\end{table}

\begin{figure}
\begin{center}
\includegraphics[width=0.5\textwidth]{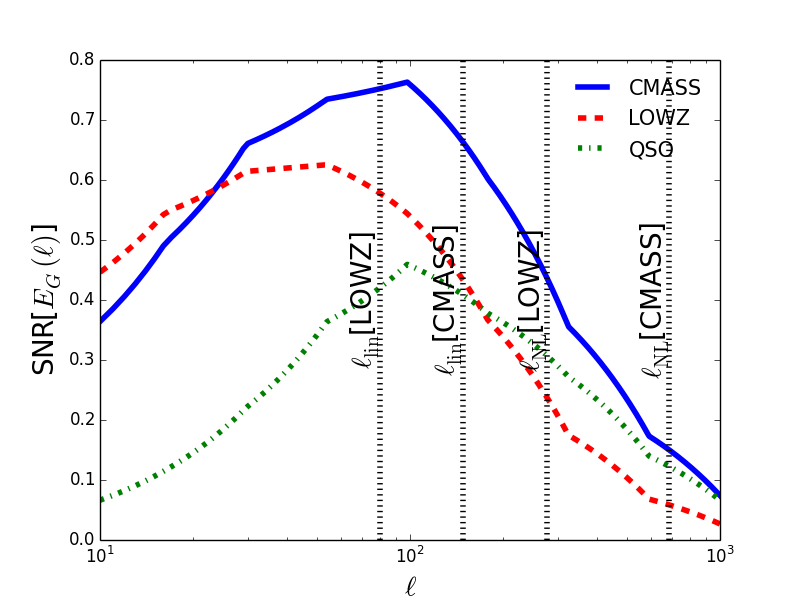}
\includegraphics[width=0.5\textwidth]{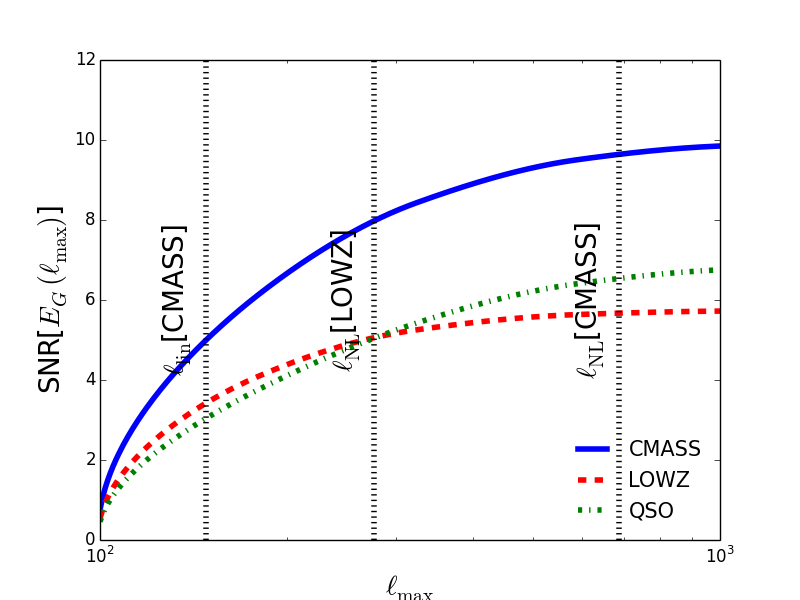}
\caption{\label{F:bossl} (\emph{Top}): The SNR of $E_G$ as a function of $\ell$ for the \emph{Planck} CMB lensing map cross-correlated with various BOSS DR11 surveys.  We also mark with vertical lines where the angular modes correspond to nonlinear scales.  We see that most of the sensitivity will come from linear scales, although our chosen cuts in $\ell$ listed in the text will lose some sensitivity. (\emph{Bottom}): The relation of the SNR of $E_G$ with an increasing $\ell_{\rm max}$ for the BOSS DR11 samples with the \emph{Planck} CMB lensing map. We see that most of the sensitivity is gained using $\ell_{\rm max}\sim300\,(500)$ for LOWZ (CMASS), while the sensitivity for the QSO sample is still increasing at $\ell_{\rm max}=1000$.}
\end{center}
\end{figure}

We plot the signal-to-noise ratio (SNR) for $10<\ell<1000$ for the CMASS, LOWZ, and quasar samples in Fig.~\ref{F:bossl}.  The peaks of all three plots vary due to the redshift of each sample.  We find that most of the signal for $E_G$ comes from linear to quasi-linear scales.  We set the maximum wavenumber within quasi-linear scales to $k_{\rm NL}$, where $\Delta(k_{\rm NL},z)=k_{\rm NL}^3P(k_{\rm NL},z)/(2\pi^2)=1$.  We find $k_{\rm NL}$[LOWZ,CMASS]=[0.354,0.466] $h$/Mpc.  For the following forecasts, we limit the angular scales used to $\ell>100$ to avoid large-scale systematic effects. In Fig.~\ref{F:bossl} we also show how the SNR increases with $\ell_{\rm max}$ for CMASS and LOWZ.  We see that most of the sensitivity is obtained by $\ell\sim500$ for CMASS and $\ell\sim300$ for LOWZ. We set these as our limits in $\ell$ for CMASS and LOWZ, while for quasars we will use all scales $\ell<1000$.

We also consider how our forecasts for $E_G$ are affected if we restrict the scales used to measure $E_G$ to only linear scales for CMASS and LOWZ.  Note we define linear scales $k<k_{\rm lin}$ as those for which the matter power spectrum $P(k)$ computed from N-body simulations differs from the linear $P(k)$ by less than a few percent.  We determine which scales are linear using the linear and $N$-body $P(k)$ predictions from Fig.~2 of \citet{2014arXiv1410.1617V}.  We find that the purely linear scales for CMASS and LOWZ are limited to $k\lesssim0.1h$/Mpc, which is significantly less than $k_{\rm NL}$ determined above for these surveys.  As can be seen in Fig.~\ref{F:bossl}, the surveys each lose about half their signal to noise if the measurements are restricted to linear scales.\footnote{$\ell_{\rm lin}$[LOWZ] $<100$, so we set $\ell_{\rm min}$[LOWZ] $=50$ when considering the total signal.}  Note that as the redshift of the survey increases, the more scales are linear.  However, $E_G$ estimates for modified gravity, particularly for $f(R)$ gravity, tend to differ from GR mainly at low redshifts.  Thus, differentiating MG models from GR requires measuring $E_G$ at quasi-linear scales.  

\begin{figure}
\begin{center}
\includegraphics[width=0.5\textwidth]{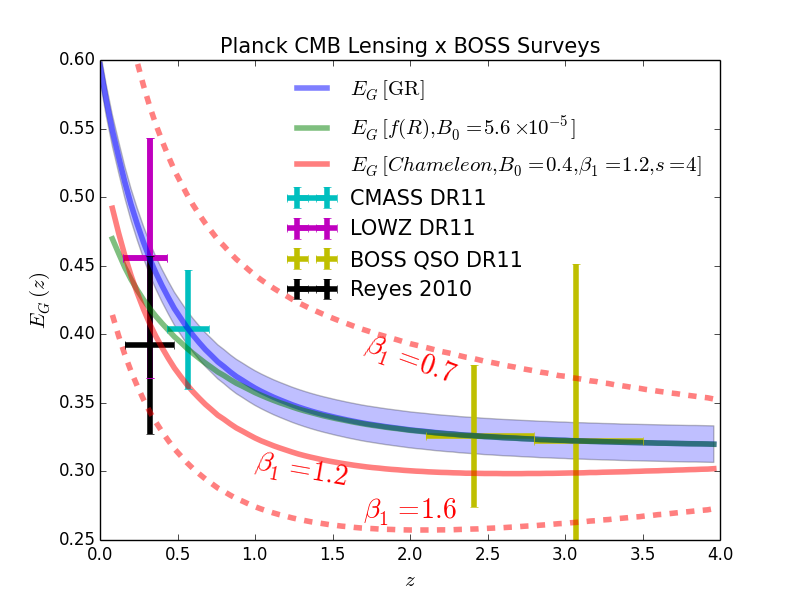}
\caption{\label{F:eglimit_boss} $E_G$ forecasts for BOSS galaxy surveys cross-correlated with the current \emph{Planck} CMB lensing map, in comparison with the latest measurement of $E_G$ using galaxy-galaxy lensing \citep{2010Natur.464..256R}.  Note that we do not translate their $E_G$ measurement from the WMAP3 cosmology \citep{2007ApJS..170..377S} to the cosmology we assume.  The band around the GR prediction corresponds to the likelihood function of $E_G$ based on \emph{Planck} and BOSS constraints on cosmological parameters.  The $E_G$ predictions for $f(R)$ gravity and chameleon gravity are averaged over the wavenumber range at every redshift corresponding to $100<\ell<500$, the range used for CMASS.  The dashed lines show chameleon gravity predictions for higher and lower values of $\beta_1$.  These surveys are not sensitive enough to tighten constraints on $f(R)$ gravity set by current measurements.}
\end{center}
\end{figure}

Next we forecast constraints on $E_G$ from BOSS surveys cross-correlated with the \emph{Planck} CMB lensing map.  We find promising results, listed in Table \ref{T:results}.  Specifically, we predict SNRs of  9.3, 5.2, and 6.8 for the CMASS, LOWZ, and QSO surveys, respectively.  We translate these into $E_G$ values in Fig.~\ref{F:eglimit_boss}.  We see that our LOWZ measurement would be comparable to that from \citet{2010Natur.464..256R}, although this may be somewhat optimistic considering LOWZ may exhibit unforeseen systematics.  However, combining all three measurements would give a SNR of 13, or an 8\% measurement.  This assumes we can measure RSD from quasars, which is very optimistic considering systematic errors that exist in quasars at large scales \citep{2013PASP..125..705P}.  We also consider our fiducial model of $f(R)$ gravity corresponding to the upper limit on the BZ parameter $B_0$.  We see that the $f(R)$ prediction differs from GR only at lower redshifts, greatly suppressing the utility of the quasar measurement and slightly increasing the utility of the LOWZ measurement.  $\chi_{\rm rms}=\sqrt{\chi^2}$ for CMASS and LOWZ are both less than unity, implying that these surveys are not able to significantly tighten constraints on $B_0$.  The sensitivity to chameleon gravity is only slightly better, in that CMASS, LOWZ, and BOSS QSOs together could differentiate models with very high (or low) values of $\beta_1$ from GR due to the rapid evolution of $E_G$ with $\beta_1$.

\subsection{Upcoming Spectroscopic Surveys} \label{S:uss}

We now consider upcoming spectroscopic surveys.  We consider two cases for the CMB lensing map, including (1) the full Planck CMB lensing map and (2) the Advanced ACTPol\footnote{private communication with Advanced ACTPol team} CMB lensing map.  In both cases, we assume the CMB lensing maps will be estimated using the temperature map and both $E$ and $B$ polarization maps, and we assume the $B$ map only contains noise.  We predict the noise in the \emph{Planck} lensing map assuming the detector sensitivity and beam sizes listed in the \emph{Planck} Bluebook \citep{2006astro.ph..4069T}.  Advanced ACTPol will survey 20,000 deg$^2$, and its increased temperature and polarization sensitivity will create a CMB lensing map that is an order of magnitude more sensitive than \emph{Planck}.  The specifications we use for Adv.~ACTPol are listed in Table \ref{T:cmbsurveys}.  For spectroscopic surveys, we consider the DESI emission line galaxy (ELG), luminous red galaxy (LRG), and quasar surveys, as well as the \emph{Euclid} H$\alpha$ survey and the WFIRST H$\alpha$ and OIII combined survey.  The properties of the surveys are listed in Table \ref{T:galsurveys}.  For DESI, we assume the same values as in the DESI Conceptual Design Report\footnote{http://desi.lbl.gov/cdr/}: $b_{\rm LRG}D(z)=1.7$, $b_{\rm ELG}D(z)=0.84$, $b_{\rm QSO}D(z)=1.2$, where $D(z)$ is the growth factor.  We also assume a 4\% error in $\beta$ within $\Delta z=0.1$ bins.  Note that Adv.~ACTPol's survey area overlaps with only $\sim$75\% of DESI's area; we take this into account in our DESI forecasts.  For \emph{Euclid} and WFIRST ELGs, we assume $b(z)=0.9+0.4z$, a fit \citep{2014PASJ...66R...1T} to semi-analytic models \citep{2010MNRAS.405.1006O} that compares well with data.  We determine the redshift distribution of \emph{Euclid} H$\alpha$ galaxies using the H$\alpha$ luminosity function from \citet{2013ApJ...779...34C} and assume a flux limit of 4$\times10^{-16}$.  This flux limit is in the middle of the range being considered, so the following \emph{Euclid} forecasts can change accordingly.  We also assume a 3\% error in $\beta$ within $\Delta z=0.1$ bins for \emph{Euclid} and WFIRST \citep{2013LRR....16....6A}.  For all subsequent forecasts, we assume $E_G$ measurements over angular scales $100\leq\ell\leq500$. 

\begin{figure}
\begin{center}
\includegraphics[width=0.5\textwidth]{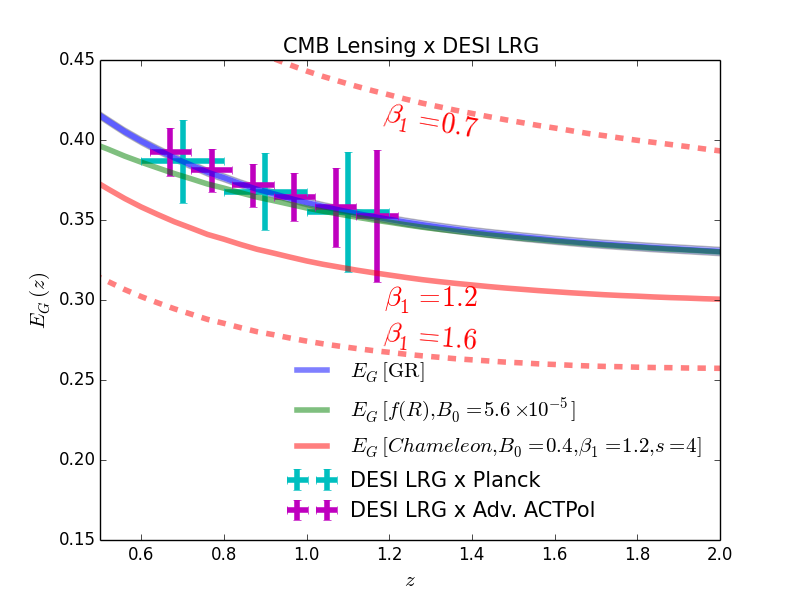}
\includegraphics[width=0.5\textwidth]{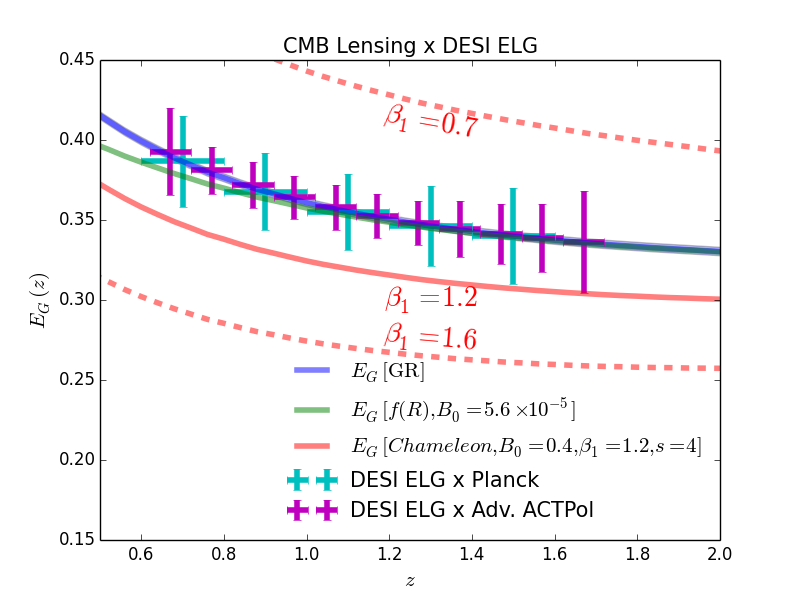}
\includegraphics[width=0.5\textwidth]{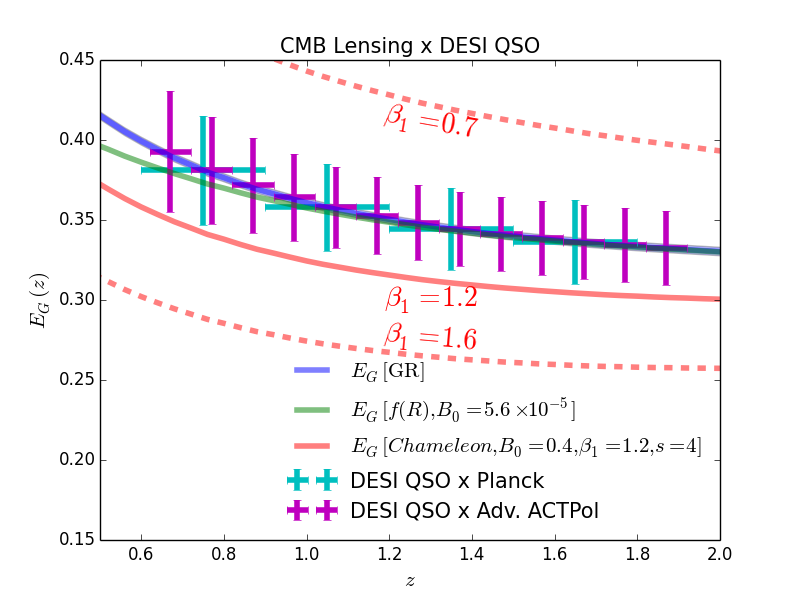}
\caption{\label{F:desi} $E_G$ forecasts for DESI galaxy surveys cross-correlated with the final \emph{Planck} CMB lensing map and with the Advanced ACTPol lensing map.  The points for Adv.~ACTPol are shifted rightward by 0.02 for clarity.  The $E_G$ predictions for $f(R)$ and chameleon gravity are averaged over the wavenumber range at every redshift corresponding to $100<\ell<500$.  The dashed lines show chameleon gravity predictions for higher and lower values of $\beta_1$.}
\end{center}
\end{figure}

\begin{figure}
\begin{center}
\includegraphics[width=0.5\textwidth]{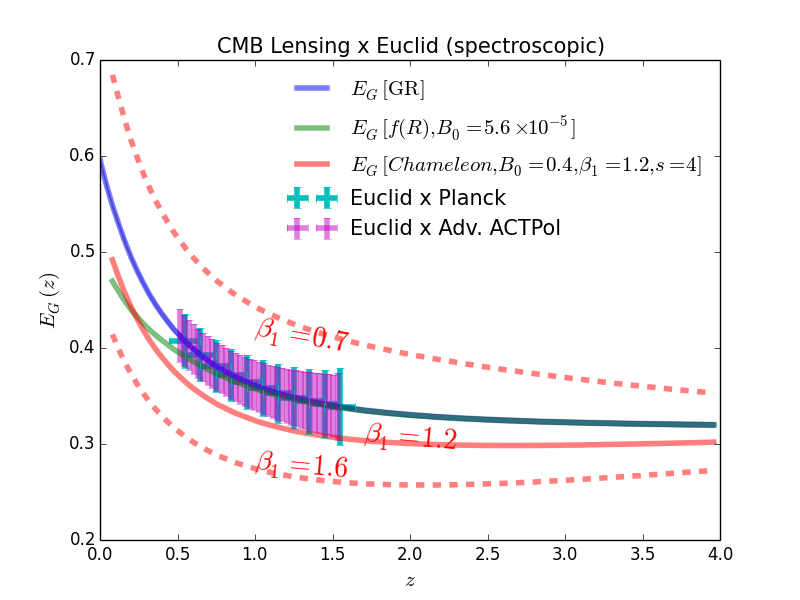}
\includegraphics[width=0.5\textwidth]{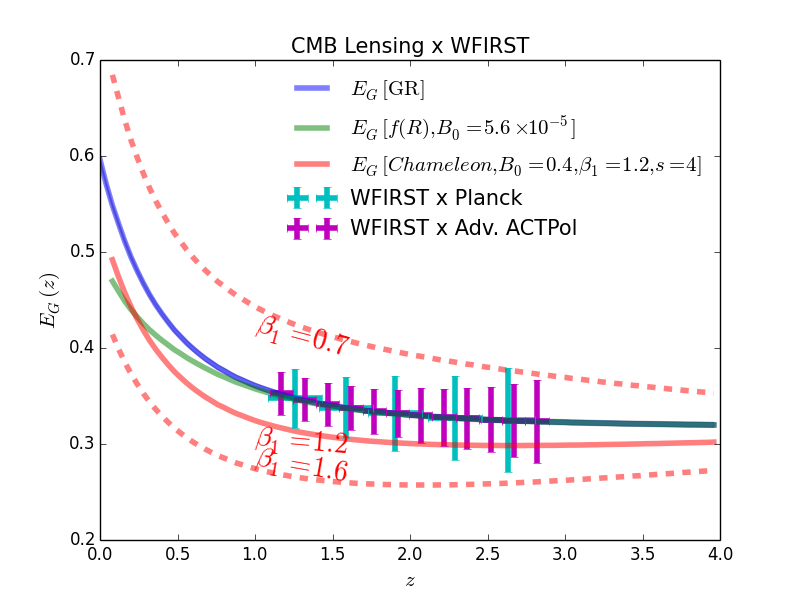}
\caption{\label{F:euclidwfirst} $E_G$ forecasts for Euclid and WFIRST galaxy surveys cross-correlated with the final \emph{Planck} CMB lensing map and with the Advanced ACTPol lensing map.  The points for WFIRST and Adv.~ACTPol are shifted rightward by 0.02 for clarity.  Note the \emph{Euclid}-Adv.~ACTPol forecasts contain 50 bins in redshift.}
\end{center}
\end{figure}

The forecasts are listed in Table \ref{T:results}, but here we list some highlights.  Figs.~\ref{F:desi} and \ref{F:euclidwfirst} show that DESI and \emph{Euclid} combined with \emph{Planck} can each measure $E_G$ almost at the 2\% level, unlike WFIRST which is limited by its small survey area.  This should allow DESI and \emph{Euclid} combined with \emph{Planck} to produce constraints of some models, and $\beta_1$ constraints should get tighter than those from BOSS.   For Adv.~ACTPol, DESI should reach a 1\% measurement of $E_G$, allowing it to differentiate GR and chameleon gravity with $\beta_1>1.1$ at the 5$\sigma$ level.  DESI produces tighter constraints than \emph{Euclid} due to its higher number density at low redshifts.  Note that we use a moderate number of redshift slices for each survey, as seen in Figs.~\ref{F:desi} and \ref{F:euclidwfirst}.  The redshift accuracy of these spectroscopic surveys would allow us to use much smaller redshift bins in an attempt to decrease errors in $E_G$.  This does not work, however, because each of these surveys are shot-noise dominated on the scales where the $E_G$ signal dominates, increasing the errors in the galaxy-CMB lensing cross-correlation.\footnote{Using the formalism in \citet{2008PhRvD..78l3506L}, the Limber approximation breaks down slightly for \emph{Euclid} with Adv.~ACTPol at some of the larger redshifts, where $\chi/\Delta\chi\gtrsim100$.  This should not affect the results significantly.}  We also considered more pessimistic errors in $\beta$, finding that increasing the error in $\beta$ by a factor of 3 did not noticeably increase $E_G$ errors from \emph{Planck}, while it increased $E_G$ errors from Adv.~ACTPol by less than 2\%.

\begin{table*}
\begin{center}
\caption{\label{T:results} Forecasts of the signal-to-noise ratio (SNR) and $\chi_{\rm rms}=\sqrt{\chi^2}$ between GR and $f(R)$ or chameleon gravity for $E_G$ measurements from various current and upcoming surveys.  For $f(R)$ gravity, we assume $B_0=5.65\times10^{-5}$.  For chameleon gravity, the first column assumes $B_0=3.2\times10^{-4}$ with $\beta_1$ and $s$ set to the base model, and the second column assumes $\beta_1=1.1$ with $B_0$ and $s$ set to the base model (see the beginning of Sec.~\ref{S:forecasts}).} 
\begin{tabular}{c|ccccc}
\hline
Survey (Galaxy $\times$ CMB lensing)&$z$&SNR&$\chi_{\rm rms}[f(R)$]&$\chi_{\rm rms}[Cham,B_0]$&$\chi_{\rm rms}[Cham,\beta_1]$\\
\hline
BOSS CMASS $\times$ \emph{Planck} (current)&0.43--0.7&9.3&0.40&0.53&0.52\\
BOSS LOWZ $\times$ \emph{Planck} (current)&0.15--0.43&5.2&0.42&0.42&0.30\\
BOSS QSOs $\times$ \emph{Planck} (current)&2.1--3.5&6.8&0.051&0.042&0.26\\
\hline
BOSS (CMASS+LOWZ+QSOs) $\times$ \emph{Planck} (current)&--&13&0.58&0.68&0.65\\
\hline
DESI ELGs $\times$ \emph{Planck} (full)&0.6--1.7&31&0.51&0.84&1.5\\
DESI LRGs $\times$ \emph{Planck} (full)&0.6--1.2&23&0.55&0.83&1.1\\
DESI QSOs $\times$ \emph{Planck} (full)&0.6--1.9&25&0.29&0.52&1.2\\
\hline
DESI (ELG+LRG+QSO) $\times$ \emph{Planck} (full)&--&46&0.80&1.3&2.2\\
\hline
DESI ELGs $\times$ Adv.~ACTPol&0.6--1.7&73&1.4&2.3&3.6\\
DESI LRGs $\times$ Adv.~ACTPol&0.6--1.2&56&1.8&2.5&2.9\\
DESI QSOs $\times$ Adv.~ACTPol&0.6--1.9&50&0.66&1.1&2.4\\
\hline
DESI (ELG+LRG+QSO) $\times$ Adv.~ACTPol&--&105&2.4&3.6&5.2\\
\hline
\emph{Euclid} (spectro) $\times$ \emph{Planck} (full)&0.5--2.0&41&0.96&1.4&2.1\\
\emph{Euclid} (spectro) $\times$ Adv.~ACTPol&0.5--2.0&83&2.4&3.2&4.1\\
\hline
WFIRST $\times$ \emph{Planck} (full)&1.05-2.9&20&0.12&0.21&0.91\\
WFIRST $\times$ Adv.~ACTPol&1.05-2.9&44&0.28&0.55&2.0\\
\hline
DES $\times$ \emph{Planck} (full)&0.0--2.0&35&1.2&1.3&1.7\\
DES $\times$ Adv.~ACTPol&0.0--2.0&78&3.0&3.3&3.9\\
\hline
LSST $\times$ \emph{Planck} (full)&0.0--2.5&84&5.1&5.2&6.0\\
LSST $\times$ Adv.~ACTPol&0.0--2.5&189&15&15&16\\
\hline
\emph{Euclid} (photo) $\times$ \emph{Planck} (full)&0.0--3.7&90&4.9&5.1&5.9\\
\emph{Euclid} (photo) $\times$ Adv.~ACTPol&0.0--3.7&205&15&15&16\\
\hline
\end{tabular}\end{center}
\end{table*}

\begin{table}
\begin{center}
\caption{\label{T:cmbsurveys} Properties of the Advanced ACTPol CMB survey.  Note that the area of the survey is 20,000 deg$^2$ and we assume $\Delta_P=\Delta_T\sqrt{2}$.} 
\begin{tabular}{c|cc}
\hline
Center Freq.&$\Delta_T$ ($\mu$K-arcmin)\footnote{per resolved pixel}&$\theta_{\rm res}$ (arcmin)\\
\hline
90 GHz&7.8&2.2\\
150 GHz&6.9&1.3\\
230 GHz&25&0.9\\
\hline
\end{tabular}\end{center}
\end{table}

\subsection{Upcoming Photometric Surveys}

In this section we consider measuring $E_G$ from upcoming photometric galaxy surveys.  These surveys, which measure less precise redshifts than spectroscopic surveys, are tailored for measuring weak lensing and not RSD.  However, the errors in $E_G$ are dominated by the CMB lensing at lower redshifts where the $E_G$ signal is highest, meaning that reducing shot noise in the lensing-galaxy cross-correlation through attaining higher number densities is be more important than having precise redshifts.  Also, upcoming photometric surveys plan to approach redshift precisions of $\sigma_z/(1+z)\sim0.05$.  Recent work has shown that upcoming photometric surveys could measure RSD \citep{2011MNRAS.415.2193R,2011MNRAS.417.2577C,2014MNRAS.445.2825A}. This may cause photometric surveys to produce competitive $E_G$ measurements.  It should be noted that Adv.~ACTPol gets close to the lensing noise limit where errors in RSD could begin to matter.  An $E_G$ measurement from a future CMB experiment that surpasses Adv.~ACTPol may reach the limit such that RSD errors may begin to dominate.  Also, the photometric redshift errors and systematic errors within photometric surveys will make measuring RSD with photometric surveys more difficult than with spectroscopic surveys \citep{2012ApJ...761...14H}.

We will construct forecasts for photometric surveys of DES, LSST, and \emph{Euclid}.  The properties of these surveys are listed in Table \ref{T:galphoto}.  For DES and LSST, we model the normalized redshift distribution in the same manner as \citet{2014JCAP...05..023F} as
\begin{eqnarray}
f_g(z)=\frac{\eta}{\Gamma\left(\frac{\alpha+1}{\eta}\right)z_0}\left(\frac{z}{z_0}\right)^\alpha\exp^{-(z/z_0)^\eta}\, ,
\end{eqnarray}
where $\alpha=1.25\,(2.0)$, $\eta=2.29\,(1.0)$, and $z_0=0.88\,(0.3)$ for DES (LSST).  For \emph{Euclid} we use estimates of the redshift distribution based on the CANDELS GOODS-S catalog \citep{2013ApJS..207...24G,2014ApJ...796...60H}.  For all three surveys we assume $b(z)=0.9+0.4z$ \citep{2014PASJ...66R...1T,2010MNRAS.405.1006O}, as in the spectroscopic case.  For RSD, we assume a 17\% error in $\beta$ over $\Delta z=0.1(1+z)$ bins for DES \citep{2011MNRAS.415.2193R}.  Since \emph{Euclid} and LSST will cover about 4 times the volume of DES, we expect the errors on $\beta$ to decrease by a factor of 2.

\begin{table}
\begin{center}
\caption{\label{T:galphoto} Properties of various galaxy photometric surveys included in our analysis.} 
\begin{tabular}{c|cccc}
\hline
Survey&$z$&Area (deg$^2$)&$N_{\rm gal}$&$\sigma_z/(1+z)$\\
\hline
DES&0.0--2.0&5000&$2.16\times10^8$&0.07\\
LSST&0.0--2.5&20,000&$3.6\times10^9$&0.05\\
\emph{Euclid}&0.0--3.7&20,000&$1.86\times10^9$&0.05\\
\hline
\end{tabular}\end{center}
\end{table}

\begin{figure}
\begin{center}
\includegraphics[width=0.5\textwidth]{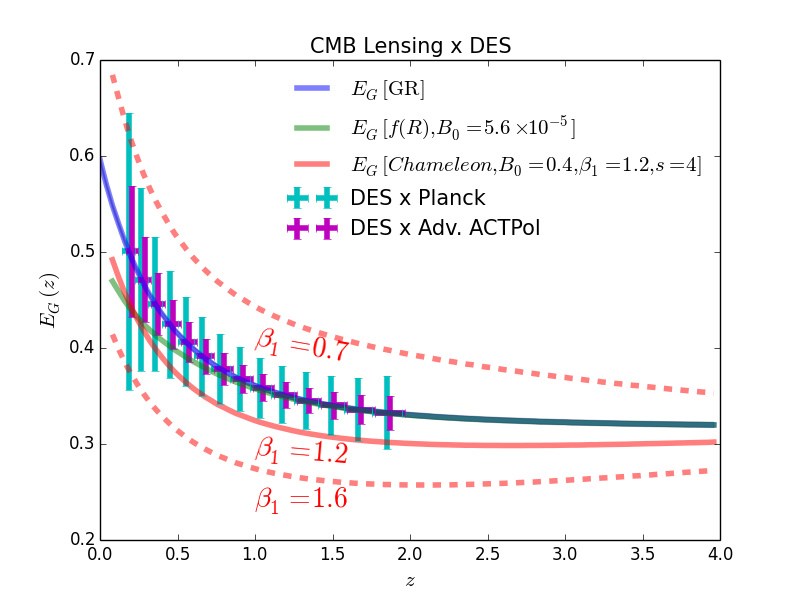}
\includegraphics[width=0.5\textwidth]{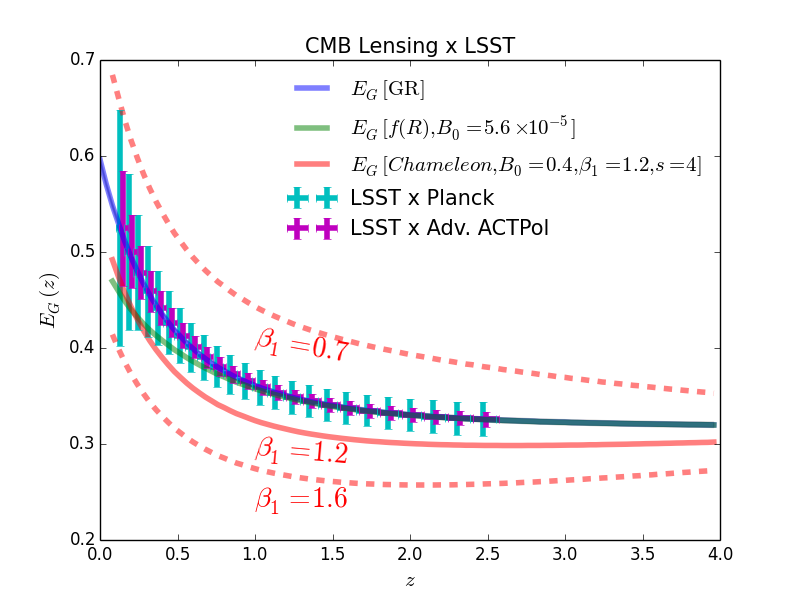}
\includegraphics[width=0.5\textwidth]{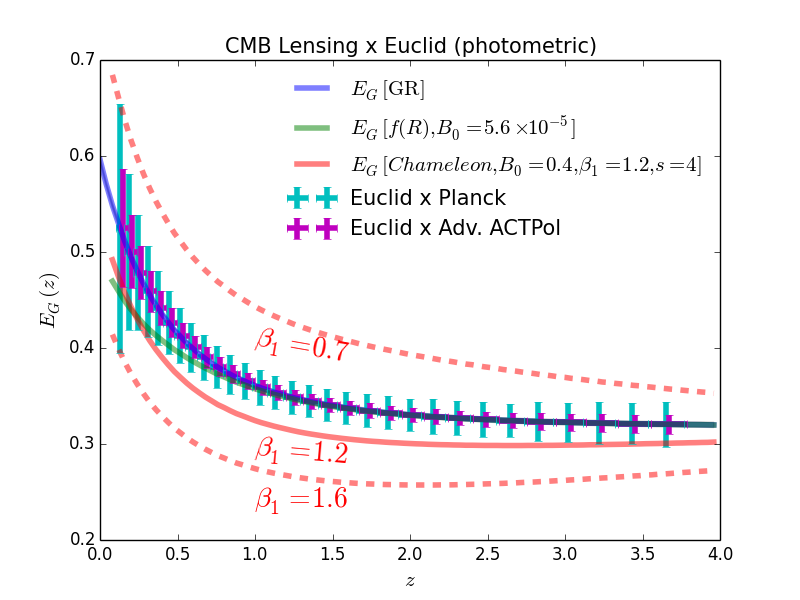}
\caption{\label{F:egphoto} $E_G$ forecasts for DES, LSST, and Euclid photometric galaxy surveys cross-correlated with the final \emph{Planck} CMB lensing map and with the Advanced ACTPol lensing map.  The points for Adv.~ACTPol are shifted rightward by 0.02 for clarity.  Note that the forecasts involving Adv.~ACTPol require a precision in the RSD parameter $\beta$ of 10\%, which may need to be obtained from a spectroscopic survey.}
\end{center}
\end{figure}

\begin{figure}
\begin{center}
\includegraphics[width=0.5\textwidth]{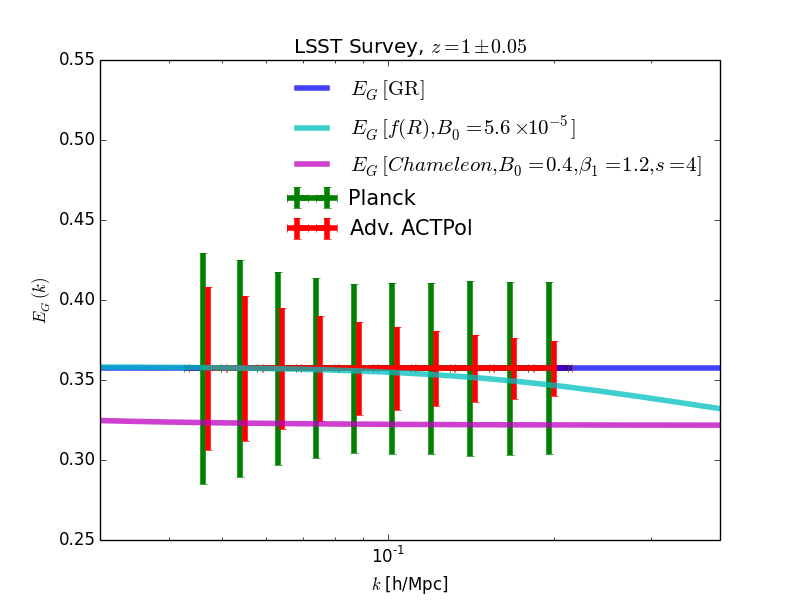}
\caption{\label{F:egkphoto} $E_G(k)$ forecasts for the LSST photometric galaxy survey in the redshift bin $z=1\pm0.05$ cross-correlated with the final \emph{Planck} CMB lensing map and with the Advanced ACTPol lensing map.  The points for Adv. ACTPol are shifted rightward by 2\% for clarity.  We also plot $E_G$ predictions for $f(R)$ gravity and chameleon gravity.}
\end{center}
\end{figure}

We find that photometric surveys can discriminate between gravity models more effectively than spectroscopic surveys, as can be seen in Fig.~\ref{F:egphoto} and Table \ref{T:results}.  DES, with its higher number density, has constraining power comparable to DESI and \emph{Euclid} with their larger survey areas.  LSST and photometric \emph{Euclid} combined with \emph{Planck} approach 1\% measurements, while substituting \emph{Planck} for Adv.~ACTPol exceeds that level.  In Fig.~\ref{F:egkphoto}, we display constraints on $E_G$ in individual $k$-bins at $z=1\pm0.05$ from LSST with both CMB surveys, finding that at this redshift the constraining power between gravity models mainly appears at smaller scales.  These two photometric surveys combined with Adv.~ACTPol could differentiate $f(R)$ gravity and GR at the 13$\sigma$ level for $B_0>10^{-7}$, severely testing $f(R)$ as a viable theory.  These surveys also would place significant tests on chameleon gravity, although the value of $\chi^2$ decreases when smaller values of $B_0$ are assumed.  Also, catastrophically increasing the errors in $\beta$ by a factor of 3 in DES increases the errors in $E_G$ using \emph{Planck} by only 5\%.  Similar increases also apply to \emph{Euclid} and WFIRST.  For Adv.~ACTPol, $E_G$ is more sensitive to this effect, in that increasing $\beta$ errors by a factor of 3 increases $E_G$ errors by 20\%.  This suggests that in this regime, RSD errors must remain low to take advantage of the power of photometric surveys to measure $E_G$, on the order of 10\% for \emph{Euclid} or LSST.

\section{Conclusions} \label{S:conclude}

In this work we consider CMB lensing as a probe of $E_G$, a statistic that differentiates between gravity models on cosmological scales. We derive $E_G$ for the general MG parametrization described by $\mu(k,z)$ and $\gamma(k,z)$, as well as for the specific MG models of $f(R)$ gravity and chameleon gravity.  We show that generally, $E_G$ for these models are scale-dependent, causing the scale-dependent $E_G$ to be useful for differentiating between MG models and GR.

We produce forecasts for current surveys, showing that BOSS spectroscopic galaxies and quasars combined with \emph{Planck} CMB lensing each measure $E_G$ at the 8\% level.  Our results suggest that CMB lensing contributes most of the error, and that measuring $E_G$ on quasi-linear scales is required to produce significant constraints.

For upcoming surveys, we find that upcoming photometric surveys will outperform spectroscopic surveys due to their higher number densities, even at the expense of having less precise redshifts.  Specifically, LSST and photometric \emph{Euclid} should produce errors on $E_G$ less than 1\%, and place very tight constraints on $f(R)$ and chameleon gravity, assuming these surveys can measure RSD with a precision of around 10\%.  However, these measurements will be limited by how well these photometric surveys can identify and remove systematic errors.  Also, since it is necessary to use $E_G$ from quasi-linear scales, measuring RSD effects at these scales will be challenging.  Finally, it is possible that more precise estimates of $E_G$ and more precise measurements of cosmological parameters may change the underlying $E_G$ errors slightly.   However, CMB lensing has the potential to probe $E_G$ with very high sensitivities without the astrophysical contaminants of galaxy-galaxy lensing, and reveal the nature of gravity.

\begin{acknowledgments}

We thank O.~Dor\'{e} and D.~Hanson for helpful comments on CMB lensing.  We also thank D.~Spergel for information on the specifications for Advanced ACTPol, as well as C.~Hirata for redshift information for the \emph{Euclid} photometric survey.  AP was supported by a McWilliams Fellowship of the Bruce and Astrid McWilliams Center for Cosmology.  SA and SH are supported by NASA grant 12-EUCLID11-0004 for this work. SH is also supported by DOE and NSF.\end{acknowledgments}

\end{document}